\documentclass{sigchi-ext}
\usepackage[T1]{fontenc}
\usepackage{textcomp}
\usepackage[scaled=.92]{helvet} 
\usepackage{graphicx} 
\usepackage{balance}  
\usepackage{booktabs} 
\usepackage{ccicons}  
\usepackage{ragged2e} 

\title{Activity and mood-based routing for autonomous vehicles}

\numberofauthors{3}
\author{%
  \alignauthor{%
    \textbf{Ankit Kariryaa}\\
    \affaddr{University of Bremen} \\
    \affaddr{Bremen, Germany} \\
    \email{kariryaa@uni-bremen.de} } \vfil \alignauthor{%
    \textbf{Tony Veale}\\
    \affaddr{University College Dublin}\\
    \affaddr{Dublin, Ireland}\\
    \email{tony.veale@ucd.ie} } \vfil \alignauthor{%
    \textbf{Johannes Sch{\"o}ning}\\
    \affaddr{University of Bremen}\\
    \affaddr{Bremen, Germany}\\
    \email{schoening@uni-bremen.de} } 
    }

\def\plaintitle{Activity and Mood-based routing for autonomous vehicles} \def\plainauthor{Ankit Kariryaa, Tony Veale, Johannes Sch{\"o}ning}
\def\plainkeywords{Routing preference; Autonomous vehicles;}

\hypersetup{%
  pdftitle={\plaintitle}, pdfauthor={\plainauthor},
  pdfkeywords={\plainkeywords}, }

\begin{document}

\maketitle

\RaggedRight{} 

\begin{abstract}
A significant amount of our daily lives is dedicated to driving, leading to an unavoidable exposure to driving-related stress. The rise of autonomous vehicles will likely lessen the extent of this stress and enhance the routine traveling experience. Yet, no matter how diverse they may be, current routing criteria are limited to considering only the passive preferences of a vehicle's users. Thus, to enhance the overall driving experience in autonomous vehicles, we advocate here for the diversification of routing criteria, by additionally emphasizing activity- and mood-based requirements. 
\end{abstract}



\section{Introduction \& Motivation}
A significant amount of our time is typically spent driving. For example, in 2014, 83.4\% of total inland travel in the EU-28 was carried out in passenger cars \cite{eurostat}. The negative effects of driving-related stress are well known ~\cite{gulian1990stress}, and while autonomous vehicles are expected to remove some of these stressors, such as e.g., the behavior of other drivers, the cognitive (over)load of driving and other ergonomic factors, other external stressors, such as e.g. environmental conditions, traffic congestion, and peripheral noise, are independent of the vehicles themselves.
To create a pleasant driving experience and to improve a user's interactions with an autonomous vehicle, we believe users should be given the possibility to limit the influence of external stressors. The related literature provides a large body of work on criteria that optimize route-finding by distance, time, safety, simplicity, efficiency and even scenery and other aesthetic preferences. We propose here the enhancement of route-finding by activity and mood to improve the user's overall experience (and enjoyment) of travel in an autonomous vehicle. 

\section{Related work}
A plethora of routing techniques supplement the most frequently used \emph{fastest} and \emph{shortest} route criteria. Based on the seminal work of Golledge in 1995 ~\cite{golledge1995defining}, Johnson et al. ~\cite{beautiful} categorizes these alternative routing techniques into the \emph{positive}, \emph{negative}, \emph{topological} and \emph{personalized}. 
The \emph{positive} category encompasses routes that are the most appealing or attractive. One such technique is described by Runge et al. ~\cite{runge2016no}. It is capable of generating scenic routes via a classification of Google Street View images. Similarly, Ali et al. ~\cite{el2013photographer} present an exploration-based route planner that learns from the routes commonly taken by photographers. As a result, it generates aesthetically-pleasing routes for city exploration and/or photography sessions. 
Routes that include the least number of unfavorable or adverse conditions to a particular user are classified as \emph{negative} routes, not because the routes themselves are negative but because routing decisions are made on the basis of discernible negatives. Shah et al. ~\cite{crowdsafe} demonstrated an algorithm for generating safe routes, deduced from the reported levels of crime  in any given locality. Li et al. ~\cite{weatherRouting} created a routing service to be used in cases of natural and man-made disasters. It first optimizes on measures of survivability, and only then on travel time.
Topological routes give higher preference to factors such as simplicity or efficiency.
Duckham and Kulik ~\cite{duckham2003simplest} proposed a routing algorithm for simple routes that are easy to describe and execute, while Ganti et al. ~\cite{GreenGPS} presented a fuel-efficient routing algorithm. Among all approaches to routing, only personalized routes take into account the preferences of individual users. 
Letchner et al. \cite{letchner2006trip} introduced the concept of an \emph{individual inefficiency ratio} (r) -- which is the degree to which a particular user deviates from the fastest route -- to fulfil personal preferences. It is based upon a large dataset of personal GPS traces. This technique suggests routes by optimizing for the preferred ratio ``r'' of the user. Delling et al. \cite{delling15} proposed a routing algorithm that creates routes based upon a user's preferences for speed level, type of road, and number and type of turns. While there has been extensive research into simple routing strategies, as well as recent research into personalized routing, we argue that the diverse needs of users cannot be satisfied with routing techniques that exploit only the passive preferences of the user. Users of non-autonomous vehicles can always optimize the suggested route based upon their current needs, but such interventions become problematic in the context of autonomous vehicles where active control is passed from the user to the vehicle itself. Hence, we propose a routing approach that considers these active elements and improves upon them.

\section{Activity and mood-based routing}
\begin{figure}
  \hspace*{-3cm} 
  \includegraphics[width=1.4\columnwidth]{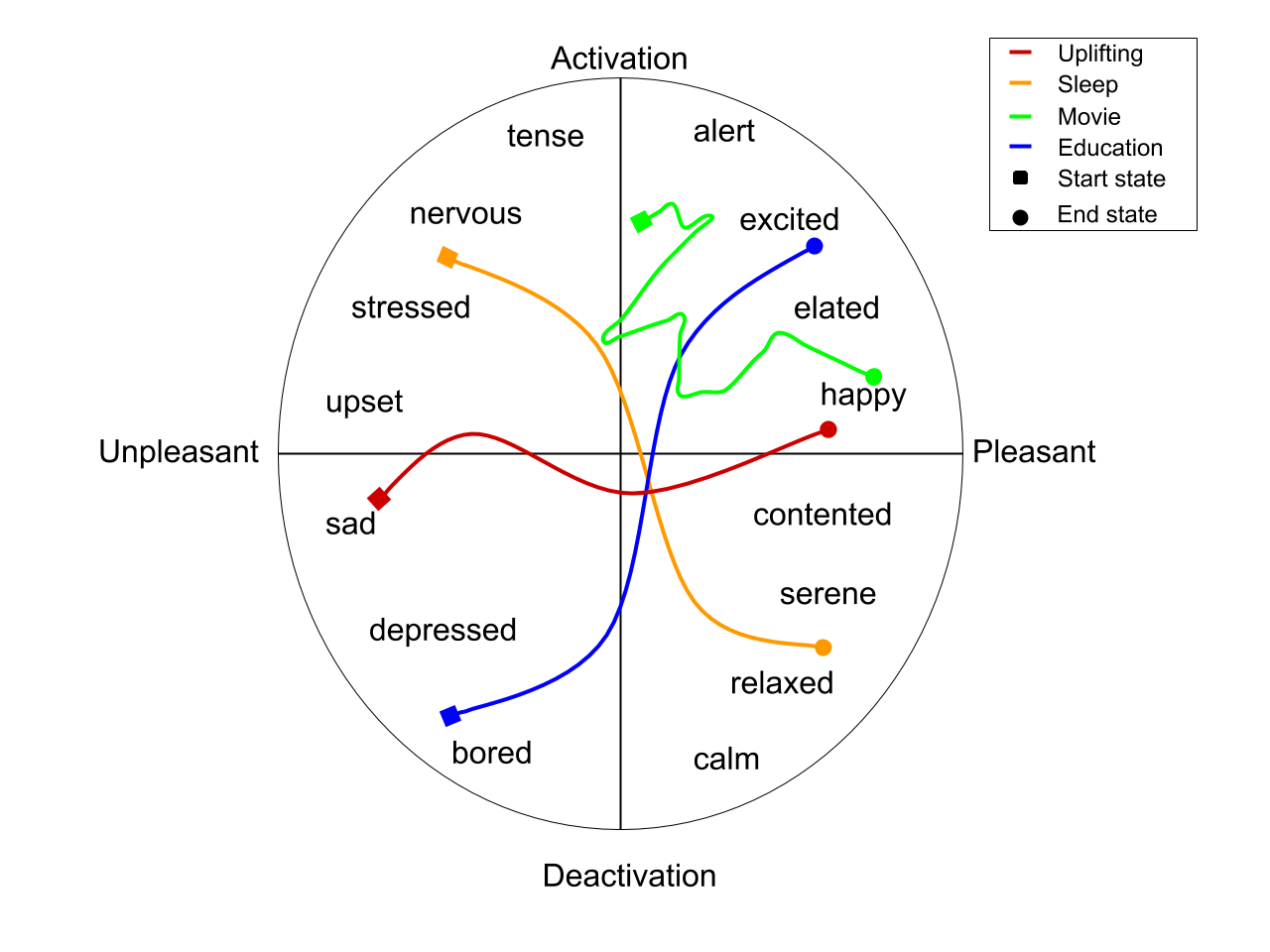}
  \caption{The route preference can be adapted to drive the emotions according to the requirement.  Based upon the graphical circumplex model of affect in ~\cite{pmid16262989}.}~\label{fig:affectModel}
\end{figure}

Research in route planning with individualized preferences shows that users prefer different routes in different conditions ~\cite{delling15, letchner2006trip}. However, little has been done to integrate the mood and activity of the user in the driving preferences, that can play a crucial role in the overall driving experience. Indeed, there is an even greater potential to incorporate the mood and activity of a non-driving user of an autonomous vehicle into routing preferences. While some of these preferences may be static -- such as avoiding areas of extreme weather or natural/social hazard -- we argue that many other preferences are dynamic, and in many cases, users can articulate their requirements explicitly (e.g., the goal of reaching a destination in the shortest time). However, in some cases a requirement cannot be mapped onto a mathematical function that can then form the basis of optimization.
The \emph{circumplex model} of affect ~\cite{pmid16262989} maps human emotions onto a valence and activation scale  (see Fig. \ref{fig:affectModel}). Since different emotions are best suited to certain activities (as was demonstrated by \cite{kapoor2001towards}), our goal for an empathic autonomous vehicle is to alter the user's mood to match the desired activity. Here, we discuss some scenarios where activity- and mood-based routing will be required.

Uplifting routes:
Picard's experiments (see ~\cite{picard1997affective}) have demonstrated the crucial role of affect in human perception and cognition, as well as its influence on such tasks as learning and decision-making. If a user experiences an unpleasant emotion while driving, one which the user would like to alter, the route itself can foster this change  (see Fig. \ref{fig:affectModel}). For instance, the route may change to visit places with associated personal value,  such as the user's favorite restaurant, lively centers of public performance, or places of natural beauty, in order to nurture more pleasant emotions. A scenic route can be chosen so as to maximize the effect of surroundings and to highlight ambient environmental conditions, such as e.g. a sunset.

Sleep-aiding routes:
If the user of an autonomous vehicle wishes to sleep during a journey, then the route preference can be optimized for the lowest level of noise and the lowest amount of other road-related disturbances, e.g. a route with the fewest turns, ensuring a smooth and continuous drive. It may also factor in additional factors such as the sleep pattern and the required sleep time of the user to promote a satisfying sleep.

Movie routes: 
If the users are watching a movie, then the route and speed of an autonomous vehicle can be varied to match the duration of the movie. It can also be chosen to suit its theme and to minimize environmental disturbances, e.g. at night a country road can be given a higher preference in order to minimize outer light disturbances. In certain cases, it may also be possible to route through locations relevant to (or analogous to) scenes in the movie, at times that match the timing of scenes in the movie, to further enhancing the experience. 

Educational routes: 
The autonomous vehicle can also propose routes based upon the educational needs of the users. For example, a route can pass through important areas and landmarks to familiarize  users with the history of an area. A route can also follow historical events, such as the course taken by an important procession or march, to give a deep insight into the locality. The route can also be chosen based upon the needs of a specific group, e.g. it can include farms, parks, museums or galleries. 

\section{Discussion and future works}
To enhance the user experience of autonomous vehicles, we have proposed new routing criteria, including both the current activity/goal and mood of the user, that should be taken into account when selecting the ``best'' route to a user's destination. These dynamic factors will diversify the route-finding process and reduce the influence of hidden externalities. 

Subsequently, an archive of chosen routes for a particular user can be used as a basis for engaging user interaction, by including ideas from both the user and the system. A creative route-finding system might come up with creative new types of route, such as routes whose road-names spell the name of a known person, or routes comprising the oldest or most historical roads, or routes whose road-names provide the answers to riddles and quiz questions. Likewise, a system may creatively choose a route based upon the plot of novel or a TV-show/movie (e.g. \emph{Game of Thrones}) that was filmed in or, set in, the locality, and entertain the user with insights about both the plot and the surrounding area. A creative system may also engage with the user in novel ways, for example, by generating a route that passes through the largest number of a certain type of building or establishment (e.g. bookshops, bars, restaurants) and play a game with users by asking them to keep track of the such establishments. In fact, a creative route-finding system should serve many of the same functions as a creative travel-companion, one that possesses much more than a map and an eye for traffic.

\section{Acknowledgements}
We would like to thank Nicolas Autzen and Tetiana Gren for valuable discussion on the topic. This work was supported by the Volkswagen Foundation through a Lichtenberg professorship.
\balance{} 

\bibliographystyle{SIGCHI-Reference-Format}
\bibliography{all}

\end{document}